\documentclass[preprint,amsmath,amssymb,aps,pre,longbibliography]{revtex4-1}

\usepackage{graphicx}
\usepackage{dcolumn}
\usepackage{bm}
\usepackage{amsmath}
\usepackage{color}
\usepackage[utf8]{inputenc}
\usepackage{cancel}
\usepackage{easyReview}
\newcommand{\figwidth}{\columnwidth}
\newcommand{\negrita}{}

\begin{document}

\title{Transport coefficients in hard-sphere fluids: thermodynamic versus kinetic descriptions}

\author{M. Hoyuelos}
\email{hoyuelos@mdp.edu.ar}
\affiliation{Instituto de Investigaciones F\'isicas de Mar del Plata (IFIMAR-CONICET), Departamento de F\'isica, Facultad de Ciencias Exactas y Naturales, Universidad Nacional de Mar del Plata, Funes 3350, 7600 Mar del Plata, Argentina}

\begin{abstract}
A thermodynamic theory for transport coefficients in hard-sphere fluids is developed from a general expression for the Onsager matrix. The theory predicts the ratio $\sigma/\sigma_{\rm id}$, where $\sigma$ is a transport coefficient and $\sigma_{\rm id}$ denotes its dilute-gas value. This ratio depends exclusively on equilibrium thermodynamic properties and can therefore be computed directly from the equation of state. Compact expressions are obtained for the thermal conductivity, viscosity, and self-diffusion coefficient. 
{\negrita These expressions agree closely with simulation data at low and intermediate densities and remain useful across the fluid range, although deviations increase near freezing, particularly for viscosity. The self-diffusion prediction substantially improves upon Enskog theory. These results support a thermodynamic description of transport coefficients relative to their dilute-gas values.}
\end{abstract}


\maketitle

\section{Introduction}

Theoretical expressions for transport coefficients are well established for dilute gases through the Chapman–Enskog theory based on the Boltzmann equation \cite{chapman}.
About a century ago, Enskog \cite{enskog} was the first to develop a transport theory that goes beyond the dilute-gas limit, using approximations valid for hard spheres. Enskog theory is a kinetic approach based on a modified Boltzmann equation that approximately holds for hard spheres at moderate densities; it includes structural information through the radial distribution function at contact. Although the theory performs poorly for self-diffusion, accurate predictions for thermal conductivity and viscosity are obtained. This was a remarkable achievement, especially taking into account that Enskog did not have the guidance of numerical simulation results. Extending these results to dense fluids with more general interaction potentials remains a longstanding challenge. {\negrita A modern account of kinetic theory, including the Boltzmann equation, the Chapman--Enskog method, transport in dilute gases, and the extension to dense hard-sphere systems through Enskog theory, can be found in Ref.\ \cite{dorfmanbook}.}

Existing approaches attempt to incorporate dense-fluid effects in various ways (see \cite{silva} for a review). Enskog-type theories incorporate refinements such as the Modified \cite{hanley} and Revised \cite{Beijeren2} Enskog theories, which improve agreement with simulations. Effective hard-sphere mappings \cite{weeks,SilvaCoelho} and free-volume theories \cite{dymond,hildebrand,batschinski,doolittle,cohen2,turnbull,macedo} capture aspects of repulsive interactions but require model-dependent prescriptions. Rosenfeld-type excess-entropy scaling \cite{rosenfeld,rosenfeld3,dyre} succeeds in organizing transport data for a broad class of fluids, yet its exponential form remains empirical and requires modifications for small density. {\negrita A central challenge is to predict the density dependence of transport coefficients from equilibrium thermodynamic properties without introducing adjustable parameters.} 

Recently, a method was proposed to relate the Onsager matrix of transport coefficients to its ideal counterpart through thermodynamic relations on a regular lattice \cite{dimuro-hoyuelos}. The ideal counterpart refers to a system where interactions can be neglected.  Denoting a transport coefficient by $\sigma$, the method implies that $\sigma/\sigma_{\rm id} = \phi$, where $\phi$ depends only on the thermodynamic state. Although the ratio $\sigma/\sigma_{\rm id}$ is not equivalent to Rosenfeld scaling, both approaches share an important feature: they link dynamic transport properties to static thermodynamic information.
Molecular dynamics simulations of pseudo-hard spheres and Lennard-Jones (LJ) systems, including noise from a Langevin thermostat, show a collapse of $\sigma/\sigma_{\rm id}$ curves as a function of particle density for different noise intensities. This result, obtained for self-diffusion \cite{marchioni} and viscosity \cite{marchioni2}, is consistent with the relation $\sigma/\sigma_{\rm id} = \phi$, since the noise does not alter the thermodynamic state. Other thermodynamics-based approaches to transport properties can be found in \cite{parrott,demirel,shapiro}.

{\negrita Here we extend the methodology of Ref.\ \cite{dimuro-hoyuelos} to fluids in continuous space and apply it to hard spheres. We partition the fluid into cells large enough to contain many particles, yet small compared with the length scales of density and temperature variations, as required by local equilibrium.}


In Sec.\ \ref{s.kvst}, an overview of the kinetic and thermodynamic approaches is presented. A unified formulation to calculate transport coefficients is developed in Sec.\ \ref{s.unified}, from which the thermal conductivity follows. The derivation of viscosity and self-diffusion coefficient requires additional hypotheses, which are presented in Sec.\ \ref{s.visco} and \ref{s.selfdif}. A comparison with numerical results to evaluate the performances of the kinetic and thermodynamic approaches is presented in Sec.\ \ref{s.comp}. Finally, the conclusions are given in Sec.\ \ref{s.conclusions}.

\section{Kinetic and thermodynamic approaches}
\label{s.kvst}

Between 1916 and 1917, Chapman and Enskog independently developed a method to obtain transport coefficients starting from the Boltzmann equation for dilute gases \cite{chapman}. The terms \textit{ideal gas}, \textit{dilute gas} and \textit{perfect gas} are used here as synonyms. More precisely, we refer to a gas that satisfies the ideal gas equation of state, has constant specific heats at constant pressure or volume, and has transport coefficients given by the Chapman-Enskog method.

The next natural step was to attempt to develop a transport theory for dense gases by extending this approach. 
Enskog proposed an extension of the Boltzmann equation for dense gases introducing a modification of the collision frequency, multiplying it by the radial distribution function at contact. As with the Boltzmann equation, the extension proposed by Enskog assumes that collisions involving three or more particles can be neglected, an assumption that is approximately valid for hard spheres at moderate densities.

Since then, many attempts have been made to extend these results to more general interaction potentials. Most of these approaches have limited predictive power because they rely on adjustable parameters. A notable exception is the Revised Enskog theory for Mie fluids presented in Ref.\ \cite{jervell}, {\negrita with reported deviations reaching 20\% in some cases.
}

{\negrita In particular, for hard spheres, the predictions of Enskog theory for thermal conductivity, $\lambda_E$, viscosity, $\eta_E$, and self-diffusion coefficient, $D_E$, can be written directly in terms of the compressibility factor $Z=P/(\rho T)$ as \cite{silva}:
\begin{align}
	\frac{\lambda_E}{\lambda_{\rm id}}
	&=b\rho\left[
	\frac{1}{Z-1}+1.2+0.755\,(Z-1)
	\right],
	\label{e.lambdaE}\\
	\frac{\eta_E}{\eta_{\rm id}}
	&=b\rho\left[
	\frac{1}{Z-1}+0.8+0.761\,(Z-1)
	\right],
	\label{e.etaE}\\
	\frac{D_E}{D_{\rm id}}
	&=\frac{b\rho}{Z-1},
	\label{e.DE}
\end{align}
where $b= 2\pi d^3/3$, $d$ is the particle diameter, and $\rho=N/V$ is the particle number density.
The dilute-gas transport coefficients are obtained from the Chapman--Enskog solution of the Boltzmann equation. Including higher-order corrections in the Sonine-polynomial expansion, they are
\begin{align}
	\lambda_{\rm id}
	&=f_\lambda\,\frac{75}{64d^2}
	\sqrt{\frac{T}{\pi m}},
	\label{e.lambdaid}\\
	\eta_{\rm id}
	&=f_\eta\,\frac{5}{16d^2}
	\sqrt{\frac{mT}{\pi}},
	\label{e.etaid}\\
	D_{\rm id}
	&=f_D\,\frac{3}{8\rho d^2}
	\sqrt{\frac{T}{\pi m}},
	\label{e.Did}
\end{align}
where, for hard spheres,
\begin{equation}\label{e.fff}
	f_\lambda=1.025,\qquad
	f_\eta=1.016,\qquad
	f_D=1.018,
\end{equation}
see Ref.\ \cite{chapman}. Here the convention $k_B=1$ is used, so that $T$ has units of energy. The factors $f_\lambda$, $f_\eta$, and $f_D$ account for higher-order terms in the Sonine-polynomial solution; setting them equal to unity gives the first Chapman--Enskog approximation. More precise values are available, for example, in Ref.\ \cite{loyalka}, but the precision of the values quoted in Eq.~\eqref{e.fff} is sufficient for the present purposes.}

For the Carnahan--Starling equation of state \cite{carnahan},
\begin{equation}\label{e.CS}
	Z=\frac{1+\varphi+\varphi^2-\varphi^3}{(1-\varphi)^3},
\end{equation}
{\negrita where $\varphi=\frac{\pi}{6}\rho d^3$ is the packing fraction. Notice that $b\rho=4\varphi$ and that $Z-1\to b\rho$ in the dilute limit, so that Eqs.~\eqref{e.lambdaE}--\eqref{e.DE} correctly reduce to the corresponding dilute-gas coefficients as $\rho\to0$.
Although more accurate equations of state are available, the Carnahan--Starling EOS remains a very accurate and convenient analytical representation throughout the stable hard-sphere fluid range. Its compressibility factor deviates from high-precision simulation results by at most about $0.3\%$ in the stable fluid phase \cite{hansengoos}. To assess the sensitivity of the transport predictions to the particular EOS, results obtained with the more accurate Kolafa--Labík--Malijevský (KLM) EOS \cite{kolafa2} are also considered in Sec.~\ref{s.comp}.
}

Now let us consider the thermodynamic approach. Naturally, equilibrium thermodynamics lacks explicit time scales, so it cannot fully determine transport coefficients on its own. The core claim, however, is that a thermodynamic transport theory can determine how transport coefficients deviate from their ideal reference values as the system departs from ideality. The ideal system is one where interactions can be neglected, the description simplifies, and transport coefficients can be calculated; in the case of a simple fluid, the ideal system corresponds to the dilute gas whose transport coefficients are obtained via the Chapman-Enskog method. 

{\negrita Enskog theory also predicts a finite bulk viscosity, which originates from collisional transfer and vanishes in the dilute-gas limit for a monatomic gas. This coefficient presents a particular difficulty for the thermodynamic approach developed here, since the present formulation predicts transport coefficients relative to their dilute-gas reference values, whereas the corresponding dilute-gas bulk viscosity is zero. We return to this limitation in the Conclusions.}

Consider a general thermodynamic description of a system consisting of a cell surrounded by a large reservoir. The system is described by extensive variables $\vec{X}$ that can be transported between neighboring cells, and entropic intensive parameters $\vec{Y}$, such that the entropy differential is $dS = \vec{Y}\cdot d\vec{X}$. The flux of $\vec{X}$ is given by $\vec{J} = - L \cdot \vec{F}$, where $L$ is the Onsager matrix, and the force is $\vec{F} = - \nabla \vec{Y}$. It was recently demonstrated \cite{dimuro-hoyuelos} that the Onsager matrix is related to its ideal counterpart, $L_{\rm id}$, through the determinant of the entropy Hessian:
\begin{equation}\label{e.L}
	L = \frac{\det H_{\rm id}}{\det H} L_{\rm id},
\end{equation}
where $H$ is the entropy Hessian matrix, $H = \frac{\partial^2 S}{\partial \vec{X}^2}$. If an extensive variable $X'$, with intensive parameter $Y'$, is not transported due to, for example, an external constraint, the Hessian of the entropy is replaced by the Hessian of the Massieu function obtained from the Legendre transform $\Phi = S - X'Y'$. The extensive variables in the ideal system must be identical to those in the real system: $\vec{X}_{\rm id} = \vec{X}$. 
It was shown that Eq.\ \eqref{e.L} reproduces known models for binary diffusion in solids and on surfaces \cite{dimuro-hoyuelos}. 

The primary purpose of this paper is to compare the predictions of Enskog theory with those of Eq.\ \eqref{e.L} applied to a hard-sphere fluid.

\section{Unified formulation}
\label{s.unified}

We wish to apply Eq.\ \eqref{e.L} to a simple fluid in continuous space. {\negrita The first part of this section holds for any interaction, but the calculation of the Hessian will be restricted to hard spheres.} Instead of extensive variables, we use the corresponding densities. The Gibbs relation is:
\begin{equation}\label{e.ds}
	ds = \frac{1}{T}\, du - \frac{\mu}{T}\, d\rho,
\end{equation}
{\negrita where $s=S/V$, $u=U/V$, and $\rho=N/V$ are, respectively, the entropy, internal-energy, and particle-number densities of a small fluid cell of volume $V$, and $\mu$ is the chemical potential per particle. The usual local-equilibrium hypothesis is assumed: the cell is small compared with the characteristic length scale of the perturbations and, at the same time, large enough to contain many particles. The comoving reference frame is used, in which the equilibrium mean velocity, $\vec v$, is zero.

To describe self-diffusion, we introduce a binary mixture obtained by assigning labels $A$ and $B$ to otherwise physically identical particles. The two species have the same mass, diameter, and interactions and differ only by their labels. Their particle densities are $\rho_A=N_A/V$ and $\rho_B=N_B/V$, with
\begin{equation}
\rho=\rho_A+\rho_B,
\qquad
\rho_D=\rho_A-\rho_B. \nonumber
\end{equation}
The variable $\rho_D$ therefore describes a local composition imbalance without introducing any physical difference between the two species.
For a binary mixture of particles $A$ and $B$, the Gibbs relation is
\begin{equation}\label{e.dsm0}
	ds = \frac{1}{T}\, du - \frac{\mu_A}{T}\, d\rho_A  - \frac{\mu_B}{T}\, d\rho_B, \nonumber
\end{equation}
where $\mu_A = \mu + T\ln(\rho_A/\rho)$ and $\mu_B = \mu + T\ln(\rho_B/\rho)$. The relation in terms of $\rho$ and $\rho_D$ is}
\begin{equation}\label{e.dsm}
	ds = \frac{1}{T}\, du - \frac{\mu_N}{T}\, d\rho  - \frac{\mu_D}{T}\, d\rho_D
\end{equation}
with
\begin{align}
	\frac{\mu_N}{T} &= \frac{\mu}{T} + \frac{1}{2}\ln\left[ (1 - r^2)/4 \right], \\
	\frac{\mu_D}{T} &= \frac{1}{2} \ln\left( \frac{1 + r}{1 - r} \right), \label{e.muD}
\end{align}
{\negrita where $\mu_N=(\mu_A+\mu_B)/2$, $\mu_D=(\mu_A-\mu_B)/2$, and $r = \rho_D/\rho$. It is assumed that the number of particles $A$ and $B$ is the same in the whole system; therefore, the equilibrium mean composition satisfies $r=0$ and $\mu_D=0$. On this equilibrium manifold, Eq.\ \eqref{e.dsm} reduces to Eq.\ \eqref{e.ds} with an additive constant in $\mu/T$ that can be ignored. Local composition fluctuations remain possible.

Thus, Eq.\ \eqref{e.ds} can be used as the description for the simple fluid and also for the binary mixture with zero equilibrium composition imbalance, in the comoving reference frame.} 

Besides thermal conductivity, we wish to obtain the viscosity, associated with the transport of the momentum density, $\vec{g} = \rho m \vec{v}$ (where $m$ is the particle mass), and the self-diffusion coefficient, associated with the transport of $\rho_D$. But neither $\vec{g}$ nor $\rho_D$ appears in Eq.\ \eqref{e.ds}. The thermodynamic space, determined by $(u,\rho)$, is two-dimensional and, moreover, $\rho$ has no associated transport coefficient since its evolution is completely given by the continuity equation. Therefore, in principle, Eq.\ \eqref{e.ds} would be useful only to obtain the thermal conductivity. Nevertheless, we propose to apply it to the other transport coefficients by introducing variables associated with the small quantities transported, as shown below. For this reason, we will not obtain an Onsager matrix where transport coefficients are simultaneously derived, since the thermodynamic space is two-dimensional, and one of the variables is the density. Instead, each coefficient will be obtained separately from different variable choices.

As a first step, we use $(T,\rho)$ as thermodynamic coordinates. Using the relation $du = \rho c_V\, dT + (\mu - T \mu_T)\,  d\rho$, where $c_V$ is the specific heat per particle at constant volume and $\mu_T = \left(\frac{\partial \mu}{\partial T}\right)_\rho$, Eq.\ \eqref{e.ds} becomes:
\begin{equation}\label{e.ds2}
	ds = \frac{\rho c_V}{T}\, dT - \mu_T\, d\rho.
\end{equation}

{\negrita We need to determine how the characteristic scales for composition and transverse-momentum transport depend on \(T\) and \(\rho\), for perturbations whose wavelength is much larger than the cell size. We consider small spatial perturbations decomposed into Fourier modes, proportional to $e^{i\mathbf{k}\cdot\mathbf{r}}$, where $\mathbf{k}$ is the wave vector. For simplicity, we focus on a single Fourier mode. For the shear-viscosity mode, the velocity perturbation is transverse to $\mathbf{k}$. Without loss of generality, $\mathbf{k}$ can be chosen along the $\hat{x}$ direction and the transverse velocity perturbation, $v_\perp$, along the $\hat{y}$ direction. The relevant transported variable is therefore the transverse momentum density $g_\perp=\rho m v_\perp$. The inclusion of momentum density among the local thermodynamic variables is standard in formulations of irreversible thermodynamics; see, for example, Sec.~6.1.2 of Ref.~\cite{lebellac}.

We introduce effective transported densities, $x_u$, $x_\perp$ and $x_D$, which quantify the small amount of the relevant extensive quantity participating in transport per unit volume. Their dependence on the background state $(\rho,T)$ is prescribed below; this dependence becomes relevant for the analysis of transport processes. We identify the fluxes of $x_u$, $x_\perp$, and $x_D$ with the physical transport fluxes of internal energy $u$, transverse momentum $g_\perp$, and composition $\rho_D$, respectively.
We define
\begin{align}
 x_u&=\epsilon'_u u_{\rm tr}=\epsilon_u\rho T,\label{e.xu}\\
 x_\perp&=\epsilon_\perp m\rho v_{\perp,\rm tr}, \label{e.xt}\\
 x_D&=\epsilon_D\rho r_{\rm tr}. \label{e.xD}
\end{align}
The subscript $\mathrm{tr}$ denotes a characteristic transport scale for the transfer of a small amount of the corresponding extensive quantity. These scales describe the dependence on the background state $(\rho,T)$, whereas the state-independent factors $\epsilon$ set the magnitude and cancel from the final transport-coefficient ratios. A physical prescription for each scale is given below.

For internal energy, we identify $u_{\mathrm{tr}}$ with the equilibrium RMS fluctuation. For a cell exchanging heat with its surroundings at fixed particle number and volume, the canonical energy variance gives
\begin{equation}
	u_{\rm tr}=u_{\rm rms}
	=\sqrt{\langle(\Delta u)^2\rangle}
	=\rho T\sqrt{\frac{c_V}{N}}.
\end{equation}
We use the same particle number when comparing the characteristic scales at different background states, so that, using $c_V=3/2$ for hard spheres, the factor $\sqrt{3/(2N)}$ can be absorbed into the state-independent coefficient $\epsilon_u$, as was already done in Eq.\ \eqref{e.xu}.

For transverse momentum and self-diffusion, we introduce additional hypotheses rather than assume the same identification. Equilibrium fluctuations of velocity and composition remain well defined even when the mean velocity and the overall composition imbalance are chosen to vanish. Their static variances, however, do not by themselves determine the dynamics of momentum or particle exchange. In particular, the composition variance depends on the arbitrary proportion of labels, although the self-diffusion coefficient does not. For randomly assigned, physically identical labels, the local composition variance conditioned on the particle number is combinatorial and independent of the interactions, whereas self-diffusion is sensitive to them.

We therefore treat $v_{\perp,\mathrm{tr}}$ and $r_{\mathrm{tr}}$ as characteristic transport scales, without identifying them with instantaneous equilibrium RMS fluctuations. Their dependence on the thermodynamic state is estimated through the mechanical-work hypotheses introduced in Sec.\ \ref{s.visco} and \ref{s.selfdif}. These hypotheses provide a constitutive closure for the proposed transport description. Any dependence on arbitrary choices of reference frame, labeling convention, or constant scale factors must cancel from the predicted transport coefficients.

Let $x$ represent any of the effective densities $x_u$, $x_\perp$, or $x_D$, and let $\tilde{x}$ denote its ideal-system counterpart. In the ideal limit, the contributions of noninteracting particles to the corresponding extensive quantity are additive and the total quantity is therefore proportional to $N$. Dividing by the volume, its density must consequently be proportional to $\rho$. Thus,
\begin{equation}\label{e.ansatzid0}
	\tilde{x}=\rho f(T).\nonumber
\end{equation}
For the three transport processes considered below, the temperature dependence can be written as
\begin{equation}
	f(T)=\epsilon T^a,\nonumber
\end{equation}
where $\epsilon$ is state independent and the constant exponent $a$ is determined separately for each case. Hence,
\begin{equation}\label{e.ansatzid}
	\tilde{x}=\epsilon T^a\rho.
\end{equation}

We write the effective density in the interacting system as
\begin{equation}\label{e.x}
	x=\tilde{x}R=\epsilon T^a\rho R,
\end{equation}
where $R$ represents the departure from the ideal reference and approaches unity in the dilute-gas limit. For hard spheres, the nonideal thermodynamic quantities entering the expressions derived below depend only on density, and therefore $R=R(\rho)$; this is a consequence of having a compressibility factor $Z$ that depends only on $\rho$. For internal-energy transport, $a=1$ and $R=1$, see Eq.\ \eqref{e.xu}.}

The procedure developed in \cite{dimuro-hoyuelos} to obtain Eq.\ \eqref{e.L} requires that the densities (or the extensive variables) in the real system be the same as in the ideal system. Therefore, we cannot apply the theory directly to the density $x$, but, instead, to its ideal limit $\tilde{x}$; we can analyze the transport properties of $\tilde{x}$ in the real system.
Differentiating \eqref{e.ansatzid}, we have:
\begin{equation}
	d\tilde{x} = \epsilon a T^{a-1} \rho \, dT + \epsilon T^a \, d\rho,
\end{equation}
and combining this with Eq.\ \eqref{e.ds2} to eliminate terms with $dT$, we obtain:
\begin{equation}
	ds = \frac{3}{2\epsilon a T^a}\, d\tilde{x} - \left(\mu_T + \frac{3}{2a}\right)\, d\rho,
\end{equation}
where the value of the specific heat for hard spheres, $c_V=3/2$, was used.

{\negrita Since the density has no independent diffusive flux in the description used here, we apply the corresponding Legendre transform, $\phi=s+\rho(\mu_T+3/(2a))$:
\begin{equation}
	d\phi = \frac{3}{2\epsilon a T^a}\, d\tilde{x} + \rho \, d\mu_T.
\end{equation}
The determinant of the Hessian of $\phi(\tilde{x},\mu_T)$ is
\begin{equation}
	\det H_\phi = - \frac{3}{\epsilon^2 T^{2a} (2\Gamma a + 3)}.
\end{equation}
where $\Gamma = \frac{\rho}{T} \left( \frac{\partial \mu}{\partial \rho} \right)_T = \left( \frac{\partial (\rho Z)}{\partial \rho} \right)_T$ is the thermodynamic factor, independent of $T$ for hard spheres, it approaches the value 1 in the ideal limit (see Appendix A).} 
Using Eq.\ \eqref{e.L}, the Onsager coefficient for $\tilde{x}$ is
\begin{equation}\label{e.LLid}
	L_{\tilde{x}} = \left(\frac{2\Gamma a + 3}{2 a + 3}\right) L_{\tilde{x}}^{\rm id}.
\end{equation}

{\negrita To obtain the last result we considered the thermodynamic properties of a cell in local equilibrium. We now consider the smooth spatial and temporal dependence of the effective densities $x$ and $\tilde{x}$, and their associated fluxes at fixed particle density. The flux of $\tilde{x}$ is:}
\begin{equation}
	\vec{J}_{\tilde{x}} =  L_{\tilde{x}} \nabla \tilde{Y} =  L_{\tilde{x}} \left(\frac{\partial \tilde{Y}}{\partial \tilde{x}}\right)_\rho \nabla \tilde{x} = - D_{\tilde{x}} \nabla \tilde{x},
\end{equation}
where $\tilde{Y} = 3/(2\epsilon a T^a)$ is the entropic intensive parameter associated with $\tilde{x}$, and $D_{\tilde{x}}$ is the diffusivity of $\tilde{x}$. {\negrita For the ideal system, the parameter $\tilde{Y}$ and the density $\tilde{x}$ are the same; then, comparing the diffusivities in the real and ideal systems, we have:}
\begin{equation}\label{e.Dxtilde}
	\frac{D_{\tilde{x}}}{D_{\tilde{x}}^{\rm id}} = \frac{L_{\tilde{x}}}{L_{\tilde{x}}^{\rm id}} = \left(\frac{2\Gamma a + 3}{2 a + 3}\right).
\end{equation}

We are actually interested in the transport of $x$, instead of $\tilde{x}$. {\negrita The flux of $x$ satisfies:}
\begin{equation}\label{e.Jx}
	\vec{J}_x = - D_x \nabla x.
\end{equation}
{\negrita From the definitions, $x=R\widetilde x$ where, as mentioned before, $R$ is a function of $\rho$. Since $\rho$ is constant, we treat $R$ as constant, in space and time. Substituting $x=R\tilde{x}$ into $\partial_t x=D_x\nabla^2x$, gives $\partial_t\tilde{x}=D_x\nabla^2\tilde{x}$. Thus,
\begin{equation}\label{e.DxDxt}
 D_x=D_{\tilde{x}}.
\end{equation}

Let $\sigma$ represent a transport coefficient: thermal conductivity $\lambda$, viscosity $\eta$, or self-diffusion coefficient $D$. With the flux identification introduced above, the physical transport coefficient $\sigma$ is defined by
\begin{equation}
	\vec{J}_x = - \sigma \nabla y
\end{equation}
where $y$ is the driving quantity: temperature for heat transport, transverse velocity for momentum transport, or composition density for self-diffusion. Also, from \eqref{e.Jx} and using that $R$ is homogeneous,
\begin{equation}
	\vec{J}_x = -D_x R\nabla \tilde{x} = - D_x R \left(\frac{\partial \tilde{x}}{\partial y}\right)_\rho\, \nabla y,
\end{equation}
Since $\tilde{x}$ is defined as the value of $x$ in the ideal system, in this case we have,
\begin{equation}
	\vec J_x^{\rm id}= -D_x^{\rm id} \nabla \tilde{x}=-D_x^{\rm id}
	\left(\frac{\partial \tilde{x}}{\partial y}\right)_\rho\nabla y.
\end{equation}
We can identify the transport coefficient as $\sigma = D_x R (\partial\tilde x/\partial y)_\rho$, and, for the ideal system, $\sigma_{\rm id}=D_x^{\rm id} (\partial\tilde x/\partial y)_\rho$. The reference dependence of $\tilde x$ on $y$ is the same in both descriptions, so the factor $(\partial\tilde x/\partial y)_\rho$ cancels from the transport coefficient ratio:
\begin{equation}\label{e.sigma0}
 \frac{\sigma}{\sigma_{\rm id}}=\frac{D_x}{D_x^{\rm id}}R.
\end{equation}
The factor $R$ is a correction arising from the definitions of the diffusivities $D_x$ and $D_x^{\rm id}$. In the real system, the diffusivity $D_x$ gives the response to the force $\nabla x$. Instead, in the ideal system, $D_x^{\rm id}$ gives the response to the force $\nabla \tilde{x}$. Different forces, related by a factor $R$, are used in each case. 
}

Using Eqs.\ \eqref{e.Dxtilde} and \eqref{e.DxDxt} in Eq.\ \eqref{e.sigma0}, we arrive at:
\begin{equation}\label{e.sigma}
	\frac{\sigma}{\sigma_{\rm id}} = \left(\frac{2\Gamma a + 3}{2 a + 3}\right) R.
\end{equation}
The result does not depend on the parameter $\epsilon$ introduced in Eq.\ \eqref{e.x}. Using this result, we immediately obtain the thermal conductivity, for which, as mentioned before, $a=1$ and $R=1$:
\begin{equation}\label{e.lambda}
	\frac{\lambda}{\lambda_{\rm id}} = \frac{2}{5}\Gamma + \frac{3}{5}.
\end{equation}

Now, the problem is to obtain $a$ and $R$ for the viscosity and the self-diffusion coefficient.

\section{Viscosity}
\label{s.visco}

{\negrita We estimate the characteristic scale of transverse collective motion through its connection with mechanical work. Define the corresponding kinetic energy per particle as \(K_{\perp,\mathrm{tr}}=m v_{\perp,\mathrm{tr}}^2/2\). A transverse velocity perturbation about a homogeneous equilibrium state satisfies \(\nabla\cdot\mathbf v_\perp=0\) (since, by definition, $\mathbf v_\perp$ is perpendicular to $\mathbf k$) and therefore does not generate a density perturbation through the linearized continuity equation. Accordingly, we consider temperature variations of $K_{\perp,\mathrm{tr}}$ at fixed $\rho$.

As a mechanical reference, consider an infinitesimal isobaric expansion. The work performed per particle is $\delta w_P=P\,d(1/\rho)$, so that
\begin{equation}
	\frac{\delta w_P}{dT}=\frac{P\alpha_P}{\rho},
\end{equation}
where $\alpha_P=-\frac{1}{\rho} \left(\frac{\partial\rho}{\partial T}\right)_P$
is the thermal expansion coefficient at constant pressure.

The quantity $\frac{\delta w_P}{dT}$ measures the expansion work associated with a unit temperature increase and thus provides a measure of the coupling between thermal changes and mechanical displacement. We assume that it also estimates the temperature sensitivity of the kinetic energy scale associated with transverse collective rearrangements:
\begin{equation}\label{e.h1}
\left(\frac{\partial K_{\perp,\mathrm{tr}}}{\partial T}\right)_\rho
\propto
\frac{\delta w_P}{dT},
\end{equation}
with a state-independent proportionality factor. The isobaric expansion is a mechanical reference process, rather than the deformation of the transverse mode itself. The derivative at fixed density compares characteristic scales at different background temperatures; it does not imply that a pure transverse mode generates a temperature perturbation at linear order. This is a heuristic relation between characteristic scales.

The equation of state $P=\rho TZ$, with $Z$ independent of $T$ for hard spheres, gives
\[
\frac{P\alpha_P}{\rho}
=\frac{Z^2}{\Gamma}
=c_P-c_V,
\]
with $c_P$ and $c_V$ the specific heats per particle at constant pressure and volume. Since this expression is independent of temperature, integration at fixed density, with $K_{\perp,\mathrm{tr}}$ vanishing as \(T\to0\), yields
\begin{equation}
	K_{\perp,\mathrm{tr}}\propto T\frac{Z^2}{\Gamma},
\end{equation}
and
\begin{equation}\label{e.vperp}
	v_{\perp,\mathrm{tr}}\propto
\sqrt{\frac{T}{m}}\frac{Z}{\sqrt{\Gamma}}.
\end{equation}
The ideal temperature dependence, $\tilde{v}_{\perp,\mathrm{tr}}\propto
\sqrt{\frac{T}{m}}$, is therefore recovered as $Z$ and $\Gamma \to1$, while the factor \(Z/\sqrt{\Gamma}\) accounts for the nonideal mechanical response.

Then, the effective transverse momentum density, Eq.\ \eqref{e.xt}, is
\begin{equation}
 x_\perp = \epsilon_\perp \rho\sqrt{mT}\frac{Z}{\sqrt\Gamma},
\end{equation}
where the proportionality constant in Eq.\ \eqref{e.vperp} has been absorbed into $\epsilon_\perp$. We can identify $a=1/2$ and $R=Z/\sqrt\Gamma$. Eq.\ \eqref{e.sigma} therefore gives
\begin{equation}\label{e.eta}
 \frac{\eta}{\eta_{\rm id}}=\frac{(\Gamma+3)}{4}\frac{Z}{\sqrt\Gamma}.
\end{equation}}

\section{Self-diffusion coefficient}
\label{s.selfdif}

{\negrita For mechanically identical particles labeled $A$ and $B$, we assign the pressure contributions $P_A=PN_A/N$ and $P_B=PN_B/N$. Consider the work per particle done by the surroundings. The work contribution assigned to population $A$, expressed per particle of the whole cell, is $\delta W_A=-P_A\,dV/N=(P_A/\rho^2)d\rho$, and similarly for $B$. The difference in the work performed on the two labeled components is
\begin{equation}\label{e.WD}
 \delta W_D=\delta W_A-\delta W_B
           =\frac{rP}{\rho^2}d\rho,
\end{equation}
where $r=(N_A-N_B)/N=\rho_D/\rho$.

When a region with a small composition imbalance expands or contracts, the two labeled populations contribute differently to the mechanical work. We assume that this difference determines the change in the characteristic chemical-potential scale associated with the diffusion event. Using the subscript $\mathrm{tr}$ for these characteristic quantities, the hypothesis is
\begin{equation}\label{e.h2}
 d\mu_{D,\rm tr}=-\delta W_{D,\rm tr},
\end{equation}
with $\delta W_{D,\rm tr}=\frac{r_{\rm tr}P}{\rho^2}d\rho$. The balance is imposed along the reference process $\mathcal P$ specified below. This is a phenomenological balance for small transported portions, applied both in the dilute limit and at finite density. It is not a thermodynamic identity for the instantaneous chemical potentials. 

For small composition differences, Eq.~\eqref{e.muD} gives $\mu_D\simeq rT$. We use the corresponding relation $\mu_{D,\rm tr}=r_{\rm tr}T$ between the characteristic quantities. Eq.\ \eqref{e.h2} becomes
\begin{equation}\label{e.muD0}
 T\left(\frac{\partial r_{\rm tr}}{\partial\rho}\right)_T d\rho
 +\left(\frac{\partial(r_{\rm tr}T)}{\partial T}\right)_\rho dT
 =-\frac{r_{\rm tr}P}{\rho^2}d\rho.
\end{equation}
In the ideal limit, $\tilde{x}_D \propto\rho\tilde r_{\rm tr}$ must preserve the linear density dependence discussed in Sec.~\ref{s.unified}. Therefore $\tilde r_{\rm tr}$ depends only on $T$, and
\begin{equation}\label{e.rt0}
 \frac{d(\tilde r_{\rm tr}T)}{dT}dT
 =-\frac{\tilde r_{\rm tr}T}{\rho}d\rho.
\end{equation}

The local process $\mathcal P$ relating $dT$ and $d\rho$ must now be specified. The Appendix B first constructs a dilute-gas reference process in which the thermal change accompanying a diffusion event occurs on its diffusion time scale, and then introduces an additional approximation to extend its local $T$-$\rho$ relation to finite density.
In this approximation,
\begin{equation}\label{e.pathD}
 \left(\frac{d\rho}{dT}\right)_{\mathcal P}\frac{T}{\rho}
 =-\frac{13}{8}.
\end{equation}
Substituting in Eq.~\eqref{e.rt0} gives
\begin{equation}
 T\frac{d\tilde r_{\rm tr}}{dT}=\frac58\tilde r_{\rm tr},
\end{equation}
whose solution is
\begin{equation}\label{e.rt}
 \tilde r_{\rm tr}=cT^{5/8},
\end{equation}
where $c$ is a state-independent constant with the appropriate units. This temperature dependence characterizes the transport scale.

In the real system, the departure from ideality is represented by $R$: $r_{\rm tr}=\tilde r_{\rm tr}R$. For hard spheres we take $R$ to depend only on density. 
Evaluating the finite-density work balance along the same reference relation between $dT$ and $d\rho$ (see Appendix B), and multiplying Eq.~\eqref{e.rt0} by $R$ gives
\begin{equation}\label{e.rt1}
 \left(\frac{\partial(r_{\rm tr}T)}{\partial T}\right)_\rho dT
 =-\frac{r_{\rm tr}T}{\rho}d\rho.
\end{equation}
Using this expression in the finite-density work balance, Eq.~\eqref{e.muD0}, and setting $P=\rho TZ$, we obtain
\begin{equation}
 T\left(\frac{\partial r_{\rm tr}}{\partial\rho}\right)_T
 -\frac{r_{\rm tr}T}{\rho}=-\frac{r_{\rm tr}TZ}{\rho}.
\end{equation}
Consequently,
\begin{equation}\label{e.lnr}
 \left(\frac{\partial\ln r_{\rm tr}}{\partial\rho}\right)_T
 =-\frac{Z-1}{\rho}.
\end{equation}
The thermodynamic identity $(\partial F/\partial V)_T=-P$ gives
\begin{equation}
 \frac{f_{\rm ex}}{T}
 =\int_0^\rho\frac{Z(\rho')-1}{\rho'}\,d\rho',
\end{equation}
where $f_{\rm ex}$ is the excess free energy per particle. Integration of Eq.~\eqref{e.lnr}, with the condition $r_{\rm tr}\to\tilde r_{\rm tr}$ as $\rho\to0$, therefore yields
\begin{equation}
 r_{\rm tr}=cT^{5/8}e^{-f_{\rm ex}/T}.
\end{equation}
The excess free energy is $f_{\rm ex} = u_{\rm ex} - Ts_{\rm ex}$, where $s_{\rm ex}$ and $u_{\rm ex}$ are the excess entropy and internal energy per particle, respectively; since, for hard spheres, $u_{\rm ex}=0$, we can replace $e^{-f_{\rm ex}/T}$ by $e^{s_{\rm ex}}$.
The effective composition density, Eq.\ \eqref{e.xD}, is
\begin{equation}
	x_D=\epsilon_D \rho T^{5/8}e^{s_{\rm ex}},
\end{equation}
where the state-independent proportionality constant $c$ was absorbed into $\epsilon_D$.
We identify $a=5/8$ and $R=e^{s_{\rm ex}}$ in Eq.~\eqref{e.sigma}. The self-diffusion coefficient is then
\begin{equation}\label{e.D}
 \frac{D}{D_{\rm id}}=\frac{(5\Gamma+12)}{17}e^{s_{\rm ex}}.
\end{equation}
The exponential factor $e^{s_{\rm ex}}$ connects this result with Rosenfeld’s excess-entropy scaling \cite{rosenfeld,rosenfeld3}, which relates reduced transport coefficients to excess entropy. Here, this factor follows from the proposed work hypothesis and the hard-sphere identity \(s_{\rm ex}=-f_{\rm ex}/T\). Equation \eqref{e.D} nevertheless differs from a simple exponential scaling law through the density-dependent prefactor \((5\Gamma+12)/17\) and the use of the dilute-gas diffusivity as the reference.}

\section{Comparison with numerical results}
\label{s.comp}

\begin{figure}
	\includegraphics[width=\figwidth]{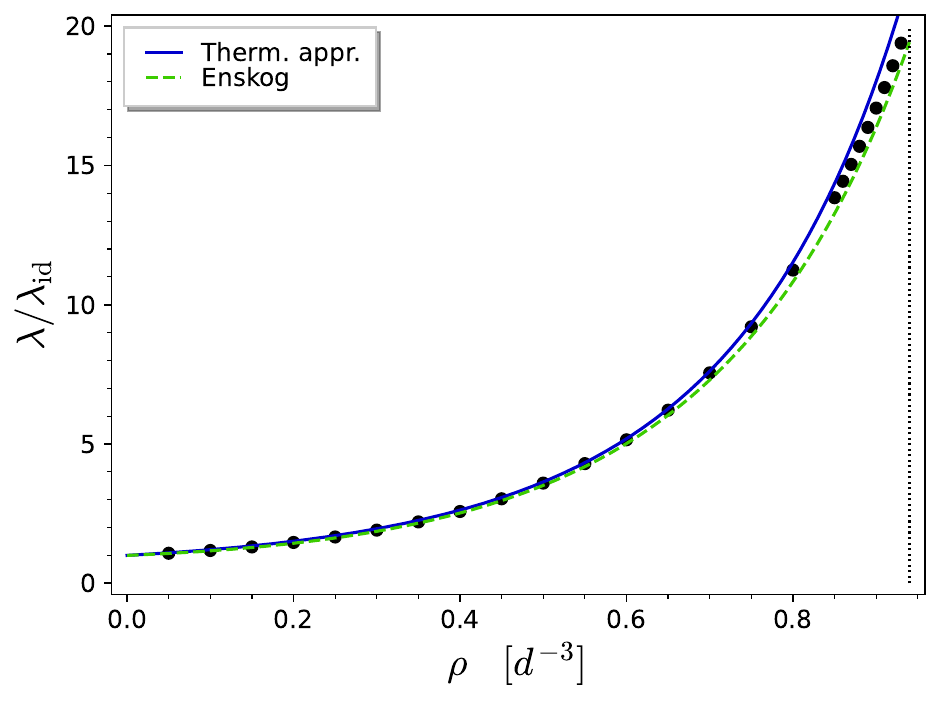}
	\includegraphics[width=\figwidth]{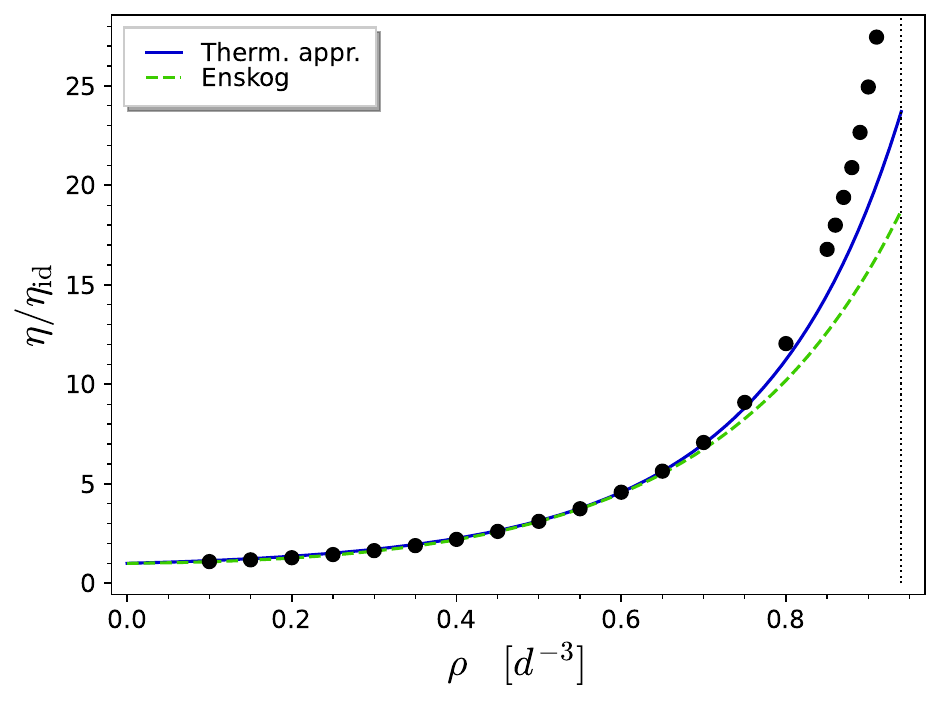}
	\includegraphics[width=\figwidth]{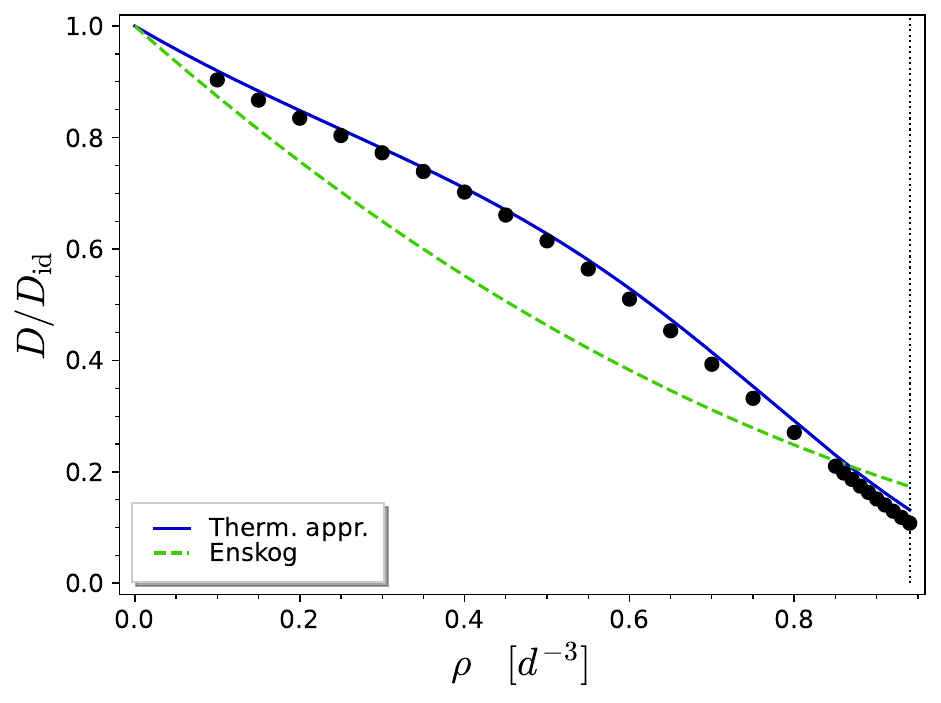}
	
	\caption{Transport coefficients normalized by their dilute-gas values as functions of particle density, $\rho$ (units $d^{-3}$, with $d$ the particle diameter), for the hard-sphere system. From top to bottom, thermal conductivity $\lambda$, viscosity $\eta$, and self-diffusion coefficient $D$. Black dots are numerical results taken from \cite{pieprzyk3} and \cite{pieprzyk2}. Continuous blue curves correspond to the present thermodynamic approach, Eqs.\ \eqref{e.lambda} for $\lambda$, \eqref{e.eta} for $\eta$, and \eqref{e.D} for $D$. Dashed green curves correspond to the kinetic approach, Eqs.\ \eqref{e.lambdaE} for $\lambda$, \eqref{e.etaE} for $\eta$, and \eqref{e.DE} for $D$.}\label{f.all}
\end{figure}

\begin{figure}
	\includegraphics[width=\figwidth]{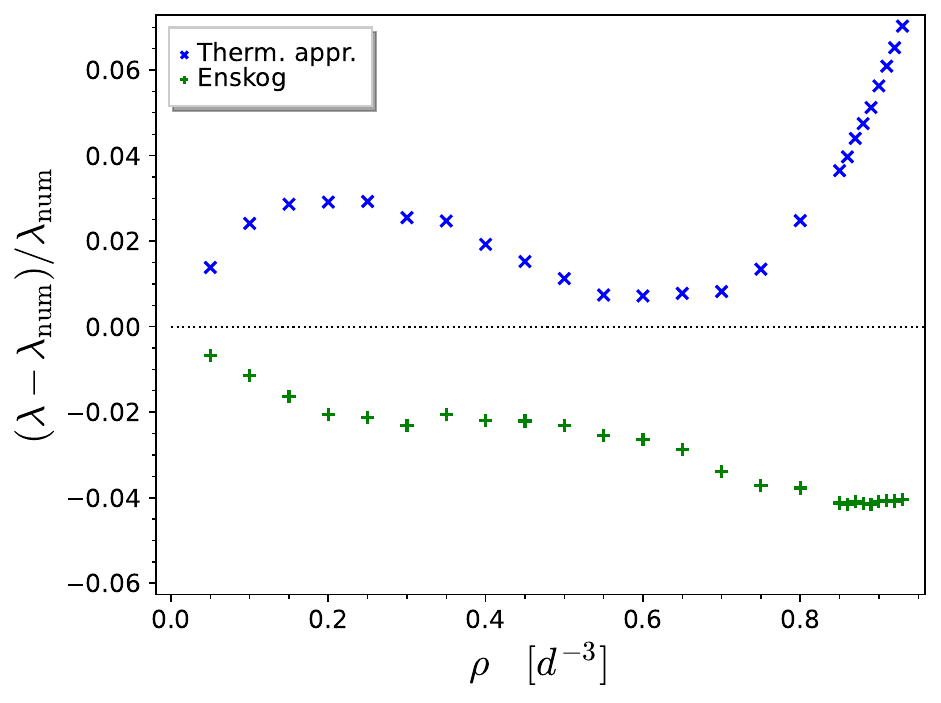}
	\includegraphics[width=\figwidth]{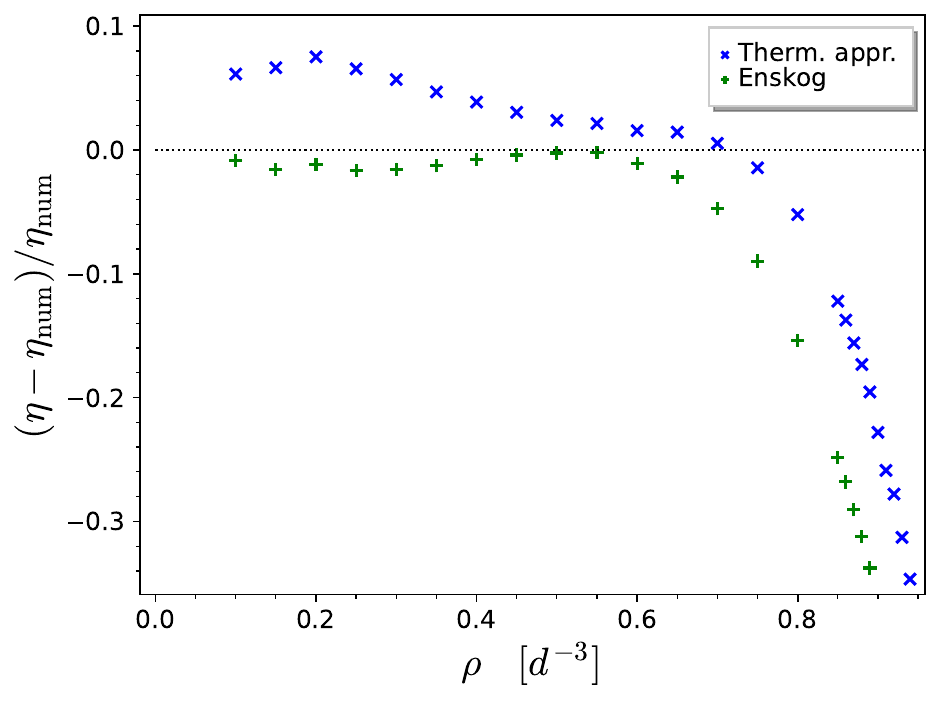}
	\includegraphics[width=\figwidth]{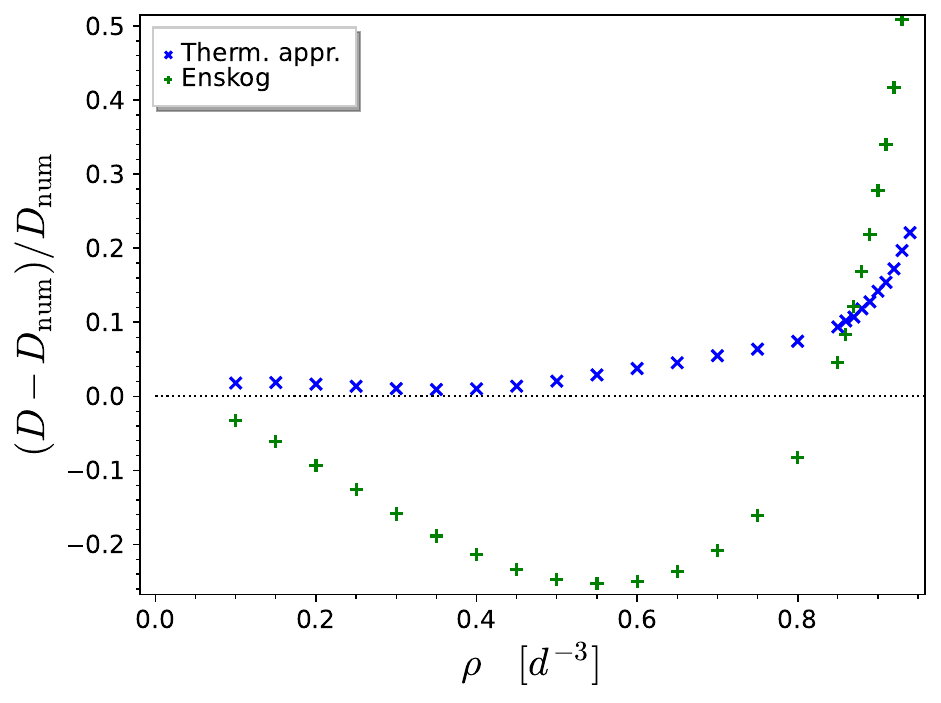}
	
	\caption{Relative deviations of the theoretical transport coefficients from simulation data, against particle density, $\rho$, for the hard-sphere system. From top to bottom, relative error of thermal conductivity, viscosity, and self-diffusion coefficient. Blue crosses  correspond to the thermodynamic approach, and green plus signs to the kinetic approach. Numerical values for $\lambda_{\rm num}$, $\eta_{\rm num}$ and $D_{\rm num}$ were taken from Refs.\ \cite{pieprzyk3} and \cite{pieprzyk2}.}\label{f.err}
\end{figure}

{\negrita The theoretical results for transport coefficients, Eqs.\ \eqref{e.lambda}, \eqref{e.eta}, and \eqref{e.D}, are compared with hard-sphere simulation data from Pieprzyk \textit{et al.} \cite{pieprzyk3} for thermal conductivity and from Pieprzyk \textit{et al.} \cite{pieprzyk2} for viscosity and self-diffusion coefficient. The Enskog predictions, Eqs. \eqref{e.lambdaE}--\eqref{e.DE}, are included for comparison. 

Figure \ref{f.all} shows the results for thermal conductivity, viscosity, and self-diffusion coefficient; the theoretical results were evaluated using the Carnahan-Starling EOS. The vertical dotted line at $\rho=0.94\, d^{-3}$ marks the freezing density \cite[p.\ 261]{frenkel}.}

Figure \ref{f.err} shows the relative error of the three transport coefficients for both approaches. These plots demonstrate that the performances of the thermodynamic and kinetic approaches are similar for thermal conductivity and viscosity, but not for the self-diffusion coefficient, where the thermodynamic approach performs better.

The performance differences can be quantified using the mean absolute percentage error (MAPE), defined for a generic transport coefficient $\sigma$ as:
\begin{equation}
	\mathrm{MAPE}_\sigma = \frac{100}{N_p} \sum_{i=1}^{N_p}\left| \frac{\sigma_i - \sigma_{i,\mathrm{num}}}{\sigma_{i,\mathrm{num}}} \right|,
\end{equation}
where $\sigma_{i,\mathrm{num}}$ is the $i$-th numerical value of the transport coefficient out of a total of $N_p$ points, and $\sigma_i$ is the theoretical prediction (from the thermodynamic or kinetic approach) at the same density as $\sigma_{i,\mathrm{num}}$.

\begin{table}[h!]
	\centering
	CS EOS
	
	\begin{tabular}{c|c|c}
		\multicolumn{3}{c}{$0\le \rho \le 0.7\, d^{-3}$} \\
		& Therm. & Kinetic \\
		\hline 
		$\lambda$ & 1.8\% & 2.2\% \\
		$\eta$ &  4.0\%
 &  1.6\%
 \\
		$D$ &  2.3\%
 &  18\%
 \\
	\end{tabular}
	\hspace{1cm}
	\begin{tabular}{c|c|c}
		\multicolumn{3}{c}{$0\le \rho \le 0.94\, d^{-3}$} \\
		& Therm. & Kinetic \\
		\hline
		$\lambda$ & 3.0\% & 3.0\% \\
		$\eta$ &  11\%
 & 16\% \\
		$D$ &  7.5\%
 &  21\%
 \\
	\end{tabular}
	\caption{Mean absolute percentage error of transport coefficients for the thermodynamic and kinetic approaches using the CS EOS. The left table corresponds to the density interval $0\le \rho \le 0.7\, d^{-3}$, while the right table covers the entire fluid range, $0\le \rho \le 0.94\, d^{-3}$.}\label{t.mape}
\end{table}

Table \ref{t.mape} presents the $\mathrm{MAPE}_\sigma$ values for the thermodynamic and kinetic approaches across two different density intervals. The left table corresponds to low and intermediate densities ($0\le \rho \le 0.7\, d^{-3}$), where both approaches exhibit their best performance. The right table covers the entire fluid density range ($0\le \rho \le 0.94\, d^{-3}$). For low and intermediate densities, both approaches offer accurate predictions for thermal conductivity and viscosity. The primary distinction arises in the self-diffusion coefficient, with {\negrita a MAPE of 2.3\% for the thermodynamic approach compared to 18\% for the kinetic approach.
} Across the entire fluid range, errors increase due to deviations near the freezing density, which are especially pronounced in the viscosity.

{\negrita 
	\begin{table}[h!]
	\centering
	KLM EOS
	
	\begin{tabular}{c|c|c}
		\multicolumn{3}{c}{$0\le \rho \le 0.7\, d^{-3}$} \\
		& Therm. & Kinetic \\
		\hline 
		$\lambda$ & 2.0\% & 2.1\% \\
		$\eta$ & 4.2\% & 1.3\% \\
		$D$ & 2.3\% & 18\% \\
	\end{tabular}
	\hspace{1cm}
	\begin{tabular}{c|c|c}
		\multicolumn{3}{c}{$0\le \rho \le 0.94\, d^{-3}$} \\
		& Therm. & Kinetic \\
		\hline
		$\lambda$ & 3.1\% & 2.9\% \\
		$\eta$ & 11\% & 16\% \\
		$D$ & 6.9\% & 21\% \\
	\end{tabular}
	\caption{Mean absolute percentage error of transport coefficients for the thermodynamic and kinetic approaches using the KLM EOS.}\label{t.eos}
\end{table}

The sensitivity of these results to the equation of state was also tested using the KLM EOS, Eq.\ (10) in Ref.\ \cite{kolafa2}. Table~\ref{t.eos} shows the MAPE values obtained with the KLM EOS. Comparing with table \ref{t.mape}, the differences are small for all three transport coefficients and over both density intervals considered. Thus, although the KLM EOS provides a more accurate description of the hard-sphere thermodynamics, the deviations observed in the transport coefficients, particularly near freezing, cannot be attributed primarily to inaccuracies of the Carnahan--Starling EOS. Since the corresponding curves are nearly indistinguishable on the scale of Fig.~\ref{f.all}, no additional plots are shown.

Accurate empirical correlations for hard-sphere self-diffusion have also been proposed by Speedy \cite{speedy} and Heyes \textit{et al.} \cite{heyes4}. These expressions provide useful representations of simulation data over the fluid range. Their purpose is, however, different from that of the present approach: they are empirical correlations constructed from transport data, whereas the present expression is obtained from equilibrium thermodynamic quantities without fitting self-diffusion data.}

\section{Conclusions}
\label{s.conclusions}

The main results of this paper are Eqs.\ \eqref{e.lambda}, \eqref{e.eta}, and \eqref{e.D}, which represent the thermal conductivity, viscosity, and self-diffusion coefficient as functions of density for the hard-sphere fluid. Compact expressions, summarized here, were obtained:
\begin{align*}
	\frac{\lambda}{\lambda_{\rm id}} &= \frac{2}{5}\Gamma + \frac{3}{5}\\
	\frac{\eta}{\eta_{\rm id}} &= \frac{(\Gamma + 3)}{4} \frac{Z}{\sqrt{\Gamma}}\\
	\frac{D}{D_{\rm id}} &= \frac{(5\Gamma + 12)}{17} \, e^{s_{\rm ex}},
\end{align*}
where the thermodynamic factor, $\Gamma$, and the {\negrita excess entropy per particle $s_{\rm ex}$}, are determined from the equation of state through the compressibility factor $Z$, Eq.\ \eqref{e.CS}. The derivation starts from the general expression for the Onsager matrix, Eq.\ \eqref{e.L}, obtained in Ref.\ \cite{dimuro-hoyuelos}. {\negrita The thermal-conductivity calculation uses the equilibrium RMS energy fluctuation to set the characteristic transport scale. For viscosity and self-diffusion, additional hypotheses specify the corresponding scales through mechanical-work relations, Eqs. \eqref{e.h1} and \eqref{e.h2}, together with the reference-process approximation described in Appendix B for the case of self-diffusion.  The proposed hypotheses relate the characteristic transverse-velocity scale to $Z$, $\Gamma$ and $T$,} and the difference in chemical potential to the difference between the mechanical work performed on the two components of a binary mixture (i.e., the work applied to the $A$ particles minus that applied to the $B$ particles). Physical arguments supporting these hypotheses were presented, and the good agreement between the resulting predictions and the numerical data provides additional evidence in their favor. Nevertheless, further verification is still required.

For the thermal conductivity and viscosity, both the Enskog kinetic theory and the thermodynamic approach provide accurate predictions, particularly at low and intermediate densities {\negrita (see Tables \ref{t.mape} and \ref{t.eos}).} {\negrita For the viscosity at high densities, however, the thermodynamic and kinetic approaches underestimate the numerical results (see Fig.\ \ref{f.all}). One possible explanation for this discrepancy, which will be investigated in future work, is that the use of the isobaric mechanical response in Eq.\ \eqref{e.h1} may become less appropriate near freezing.}

For the self-diffusion coefficient, the thermodynamic approach yields a remarkably higher accuracy than the kinetic theory. It is well known that the latter fails to account for the dynamic memory effect responsible for the algebraic long-time decay of the velocity autocorrelation function, instead of the exponential decay predicted by kinetic theory. As a consequence, the Enskog theory underestimates the self-diffusion coefficient at low and intermediate densities. The improved self-diffusion prediction suggests that the thermodynamic description accounts effectively for density-dependent contributions missed by Enskog theory.

{\negrita A limitation of the present formulation concerns the bulk viscosity. The transport coefficients considered here are obtained relative to their dilute-gas values, whereas the bulk viscosity of a dilute monatomic gas vanishes. Consequently, the present ideal-gas reference cannot be used to obtain a finite bulk viscosity. The general thermodynamic approach, however, need not necessarily be restricted to an ideal reference system. A possible extension would be to identify a nonideal reference system for which the bulk viscosity is finite and independently known. Whether a useful reference system of this kind exists remains an open question for future work.

These results support the use of equilibrium thermodynamic properties to describe transport coefficients relative to their dilute-gas values. The procedure is based on the general expression for the Onsager matrix derived in Ref.\ \cite{dimuro-hoyuelos}. Extending the analysis to fluids with more general interaction potentials is a natural next step.}

\begin{acknowledgments}
{\negrita 	I thank M. A. Di Muro, M. Sampayo Puelles, L. Marchioni, P. Giménez, and A. Alés for helpful discussions. This work was partially supported by the Universidad Nacional de Mar del Plata (UNMdP, Argentina, Grant No. 80020250500008MP).}
\end{acknowledgments}

\section*{Appendix A}
\label{app.hessian}

{\negrita For completeness, we give here the explicit Hessian matrix used in Sec.~\ref{s.unified}. The natural variables of the Massieu function are $(\tilde{x},\mu_T)$, and
\begin{equation}
	d\phi=\tilde{Y}\,d\tilde{x}+\rho\,d\mu_T,
	\qquad
	\tilde{Y}=\frac{3}{2\epsilon aT^a}.
\end{equation}
The Hessian of $\phi$ is
\begin{equation}
	H_\phi = \frac{\partial^2 \phi}{\partial (\tilde{x},\mu_T)^2} = \frac
	{\partial(\tilde{Y},\rho)}{\partial(\tilde{x},\mu_T)}.
\end{equation}
It is convenient to write the inverse Hessian,
\begin{equation}
	H_\phi^{-1}
	=
	\frac{\partial(\tilde{x},\mu_T)}
	{\partial(\tilde{Y},\rho)}.
\end{equation}

From $\tilde{x}=\epsilon T^a\rho$ and the thermodynamic relations valid for hard spheres
\begin{equation}
	\left(\frac{\partial\mu_T}{\partial T}\right)_\rho
	=-\frac{3}{2T},
	\qquad
	\left(\frac{\partial\mu_T}{\partial\rho}\right)_T
	=\frac{\Gamma}{\rho},
\end{equation}
we have
\begin{equation}
	H_\phi^{-1}
	=
	\begin{pmatrix}
		-\dfrac{2}{3}a\epsilon^2\rho T^{2a}
		&
		\epsilon T^a
		\\[7pt]
		\epsilon T^a
		&
		\dfrac{\Gamma}{\rho}
	\end{pmatrix}.
	\label{e.Hinv}
\end{equation}
Its determinant is
\begin{equation}
	\det H_\phi^{-1}
	=
	-\frac{\epsilon^2T^{2a}}{3}
	\left(2a\Gamma+3\right),
\end{equation}
and therefore
\begin{equation}
	\det H_\phi
	=
	-\frac{3}
	{\epsilon^2T^{2a}(2a\Gamma+3)}.
	\label{e.detHphi}
\end{equation}

For the ideal system, $\Gamma=1$, so that
\begin{equation}
	\det H_\phi^{\rm id}
	=
	-\frac{3}
	{\epsilon^2T^{2a}(2a+3)}.
\end{equation}
Consequently,
\begin{equation}
	\frac{\det H_\phi^{\rm id}}
	{\det H_\phi}
	=
	\frac{2a\Gamma+3}{2a+3},
\end{equation}
which leads directly to Eq.~\eqref{e.LLid} through Eq.~\eqref{e.L}.}

\section*{Appendix B}

{\negrita In this appendix, we give a heuristic estimate of $(d\rho/dT)_{\mathcal P}$ for the local process accompanying a diffusion event in a dilute gas. Consider a small cell with fixed total particle number that exchanges heat and work with the surrounding fluid. Its temperature $T$ and density $\rho$ undergo small changes about equilibrium. For a spatial scale $k^{-1}$, the diffusion time is $\tau_D=(D_{\rm id}k^2)^{-1}$. We assume that, for this process, the accompanying thermal change occurs on this time scale:
\begin{equation}
 \frac{dT}{dt}=-D_{\rm id}k^2(T-T_0),
\end{equation}
where $T_0$ is the equilibrium temperature. Approximating the heat exchange by Fourier conduction at the same spatial scale, the rate of heat transfer per particle is
\begin{equation}
 \frac{dq}{dt}=-\frac{\lambda_{\rm id}}{\rho}k^2(T-T_0).
\end{equation}
The coefficients are evaluated at equilibrium, as appropriate for small departures from equilibrium. Eliminating time gives
\begin{equation}\label{e.dtdr0}
 \delta q=\frac{\lambda_{\rm id}}{\rho D_{\rm id}}dT.
\end{equation}
Here $q$ denotes the heat transferred per particle, and $\delta q$ is its inexact increment. The first law in the ideal reference, including the compression work received by the cell, is
\begin{equation}\label{e.linid}
 c_VdT=\delta q+\frac{T}{\rho}d\rho.
\end{equation}
Combining these expressions gives
\begin{equation}
 \left(\frac{d\rho}{dT}\right)_{\mathcal P}\frac{T}{\rho}
 =c_V-\frac{\lambda_{\rm id}}{\rho D_{\rm id}}.
\end{equation}
For hard spheres, $c_V=3/2$ and, using the first order Chapman--Enskog coefficients, we have $D_{\rm id}=8\lambda_{\rm id}/(25\rho)$. Therefore,
\begin{equation}
 \left(\frac{d\rho}{dT}\right)_{\mathcal P}\frac{T}{\rho}
 =-\frac{13}{8}.
\end{equation}

To extend the reference process to finite density, we assume that interactions modify the local rates of temperature and density change by a common positive factor $M$, evaluated at the background equilibrium state $(T_0,\rho_0)$:
\begin{equation}
\left(\frac{\dot T}{T_0}\right)_{\rm real}
=M\left(\frac{\dot T}{T_0}\right)_{\rm id},
\qquad
\left(\frac{\dot\rho}{\rho_0}\right)_{\rm real}
=M\left(\frac{\dot\rho}{\rho_0}\right)_{\rm id}.
\end{equation}
The physical motivation is that the surrounding particles may delay or facilitate a local rearrangement while leaving approximately unchanged the ratio between the temperature and density changes it produces. Interactions would then primarily affect how fast the process occurs, rather than its local trajectory in the $T$-$\rho$ plane. The common factor cancels from the ratio, giving
\begin{equation}
	\left(\frac{d\ln \rho}{d\ln T}\right)_{\mathcal{P},\rm real} = 	\left(\frac{d\ln \rho}{d\ln T}\right)_{\mathcal{P},\rm id} = -\frac{13}{8}.
\end{equation}
}

\bibliography{tttHS.bib}

\end{document}